\documentstyle[12pt]{article}

\input epsf

\def\bild#1#2{    
        \vspace*{-5mm}
        \begin{center}
        \begin{math}
        \epsfxsize#2cm
        \epsffile{#1}
        \end{math}
        \end{center}
        }

\newcommand{\ppp}[1]{%
        \setbox0=\hbox{#1}%
        \kern-.02em\copy0\kern-\wd0
        \kern+.04em\copy0\kern-\wd0
        \kern-.02em\raise.0217em\box0}

\newcommand{\lsim}{
 \mathrel{\setbox0=\hbox{$<$}\raise0.6ex\copy0\kern-\wd0
 \lower0.65ex\hbox{$\sim$}}}

\newcommand{\gsim}{
 \mathrel{\setbox0=\hbox{$>$}\raise0.6ex\copy0\kern-\wd0
 \lower0.65ex\hbox{$\sim$}}}

\begin{document}
%
\begin{titlepage}
\renewcommand{\thefootnote}{\fnsymbol{footnote}}
\makebox[2cm]{}\\[-1in]
\begin{flushright}
\begin{tabular}{l}
TUM/T39-97-19
\end{tabular}
\end{flushright}
\vskip0.4cm
\begin{center}
  {\Large\bf
     Gluon Polarization in  Nucleons\footnote{Work supported in
    part by BMBF}
 }\\ 

\vspace{2cm}

A. Saalfeld, G. Piller and 
L.\ Mankiewicz\footnote{On leave of absence from N. Copernicus
Astronomical Center, Polish Academy of Science, ul. Bartycka 18,
PL--00-716 Warsaw (Poland)} 

\vspace{1.5cm}

{\em Institut f\"ur Theoretische Physik, TU M\"unchen, Germany}

\vspace{1cm}

{\em \today}

\vspace{3cm}

{\bf Abstract:\\[5pt]} \parbox[t]{\textwidth}{ 
  In QCD the gauge-invariant gluon
  polarization $\Delta G$ in a nucleon can be defined either in a
  non-local way as the integral over the Ioffe-time distribution of polarized
  gluons, or in light-cone gauge as the forward matrix element of the 
  local topological current. We have
  investigated both possibilities within the framework of QCD sum rules.
  Although the topological current  is built from local
  fields, we have found that its matrix element retains sensitivity to large 
  longitudinal distances.
  Because QCD sum rules produce
  artificial oscillations of the Ioffe-time distribution of polarized 
  glue at moderate and large
  light-like distances, the calculation of the matrix element of the 
  topological
  current results in a small value of $\Delta G(\mu^2 \sim 1\, {\rm GeV}^2)
  \approx  0.6 \pm 0.2$.  
  In a more consistent approach QCD sum rules are used
  to describe the polarized gluon distribution only at small light-like
  distances.  
  Assuming that significant contributions to $\Delta G$ arise only
  from longitudinal length scales not larger than the nucleon size 
  leads to $\Delta G(\mu^2 \sim 1\, {\rm GeV}^2)
  \approx 2 \pm 1$.  }

\end{center}
\end{titlepage}
\setcounter{footnote}{0}

\newpage

\section{{\bf Introduction}}

\vskip 0.5 cm

Thanks to celebrated factorization theorems \cite{Fact} hard scattering of
highly virtual probes from nucleons can be characterized in QCD by universal,
process independent, non-perturbative distribution functions which contain all
relevant information about the long-distance dynamics of the target. 
At twist-$2$ these are identified in the framework of the
QCD-improved parton model with scale-dependent quark and gluon 
light-cone distributions. 
For spin-${1}/{2}$ targets they 
include unpolarized quark and gluon distributions and their
polarized counterparts related to longitudinal and transverse target
polarizations. 
Quantum numbers and chiral properties of the hard probe determine 
which particular set of distribution functions can be accessed
in a specific process.
Until now most information about the nucleon structure has been
obtained from deep-inelastic lepton scattering experiments.  In particular,
recent measurements of polarized deep-inelastic scattering at CERN \cite{CERN}
and SLAC \cite{SLAC} have shown that a relatively small fraction of the nucleon
spin is carried by quarks.  This has started an ongoing debate about remaining
contributions to the nucleon spin \cite{Deb96} which may result from gluon
polarization and orbital angular momentum.  Here especially the polarized gluon
distribution $\Delta G(u,\mu^2)$\footnote{Throughout this paper we
denote the Bjorken variable $x_{\rm Bj}$ by $u$ to avoid confusion with
the space-time variable $x$.}  
became of interest since it turned out to be
measurable in future high-energy experiments, e.g. charm and direct photon
production \cite{COMPASS,RHIC}.

At present the only available information about
the magnitude of the total gluon polarization
\begin{equation}
\Delta G(\mu^2) = \int_0^1 du \,\Delta G(u,\mu^2)
\end{equation}
results from the analysis of scaling violations in the polarized structure
function $g_1(u,Q^2)$.  Here a relatively large gluon polarization 
has been found
even at a low normalization scale $\Delta G(\mu^2 = 1 {\rm GeV}^2) = 1.6 \pm
0.9$ \cite{Alt96}.  The main objective of the present paper is to discuss a
framework in which one may determine $\Delta G(\mu^2)$ using presently known
non-perturbative methods.

The link between parton model ideas and QCD is provided by the 
Operator Product Expansion (OPE) \cite{Fact}.  
It allows to relate moments of parton distributions to matrix 
elements of local operators 
with appropriate quantum numbers\footnote{We 
  define the $n$-th moment of a distribution $F(u)$  
  as $\Gamma_n[F] = \int_0^1 du u^{n-1} F(u)$.}. 
The latter are computable either in some
approximate way using e.g. model descriptions of hadronic structure 
\cite{Bag,Inst} or QCD sum rules \cite{QCDSR}, or at least in principle, 
through lattice simulations  \cite{Latt}.

An alternative, but completely equivalent picture views twist-$2$ parton
distribution functions as normalized Fourier transforms of nucleon matrix
elements of non-local QCD operators \cite{Collins}, constructed as 
gauge-invariant overlap of two quark or gluon fields separated by a light-like
distance. 
Since the  first few coefficients of the Taylor expansions of these  
non-local matrix elements around the origin are given by matrix
elements of local, low-dimensional twist-$2$ operators, the domain of small
longitudinal distances is determined by known non-perturbative QCD methods. 
However for increasing longitudinal distances more and more twist-$2$ matrix
elements are required and soon one enters a region where lattice QCD and other
non-perturbative approaches are practically not applicable any more.
Phenomenologically, for unpolarized charge conjugation even quark
and gluon distributions one observes a transition between the region 
of small and large longitudinal distances around a length scale which
corresponds to the electromagnetic nucleon size of around $2 \,{\rm fm}$ 
\cite{Weig96}.
Beyond that scale the matrix elements of the corresponding string operators 
become smooth and approximately flat as functions
of longitudinal distance --  at least at low normalization scales. 
The matrix elements of the string operators which correspond to 
$C$-odd unpolarized quark distributions decrease and become small 
at longitudinal distances beyond $2\,{\rm fm}$. 
In both cases, however, the region of large longitudinal distances
is beyond the scope of presently available non-perturbative methods,  
and one has to resort to approximations such as Regge theory.  
An ideal non-perturbative QCD observable should therefore avoid 
contributions from large longitudinal distances, 
i.e.  it should be sensitive only to those degrees of freedom which reveal 
themselves at length scales which are not larger than
the nucleon size.

Although the total gluon polarization receives most interest from 
a phenomenological point of view, 
establishing a framework for its evaluation 
has its own importance.
This is due to the fact that there is no local, gauge 
invariant operator which  can serve as a ``gluon polarization partonometer'', 
i.e. which yields a matrix element associated with 
$\Delta G(\mu^2)$. 
In general two different ways to define $\Delta G(\mu^2)$  have been 
discussed in the literature so far. 
It has been known for a long time that the gluon polarization 
can be related to the matrix element 
of a gauge invariant, but non-local gluonic string operator 
\cite{Man90,BB91}:
\begin{equation}
O_{\rm NL} = n_\mu n_\nu \int_0^\infty d \lambda \, 
{\mbox Tr}\, G^{\mu\xi}(\lambda n)[\lambda n;0]
{\tilde G}_\xi\,^\nu(0). 
\label{ONL}
\end{equation}
Here $G^{\mu\nu}$ and ${\tilde G}^{\mu\nu}$ denote the gluon field 
strength and its dual, respectively, 
and $n_\mu$ is a light-like vector with $n^2 = 0$, $n \cdot a = a^0
+ a^3 \equiv a^+$ for any four-vector $a$. 
The trace in (\ref{ONL}) is  performed in color space and 
the  path-ordered exponential in the adjoint
representation $[\lambda n;0]$ guarantees gauge invariance.
Note however that so far most experience  has been obtained in 
dealing with matrix elements of local operators, and
therefore it is not clear a'priori how one should 
apply the non-local operator (\ref{ONL}) in practice.  
In our  recent work \cite{MPS97} we have argued that a computation of $\Delta
G(\mu^2)$ using the operator $O_{\rm NL}$ at some low normalization scale is
possible, but requires an insight into the nature of contributions arising from
large longitudinal distances. 
Once one accepts a point of view,  supported e.g. by  Regge theory, that 
despite its non-local character the gluon polarization 
$\Delta G(\mu^2)$ receives only minor contributions from large longitudinal 
distances, an approximate ``gluon polarization partonometer''  
can be constructed in a gauge-invariant way. 
 In its simplest version it takes into
account information encoded in only two first computable QCD moments of the
polarized gluon distribution function.

On the other hand in  light-cone gauge $n\cdot A = A^+ = 0$ the operator 
$O_{\rm NL}$ assumes a local form, identical to the 
$n \cdot K = K^+$ component of the topological current 
\cite{Man90,BB91}:
\begin{equation}
K_\mu = \frac{\alpha_s}{2 \pi}
\epsilon_{\mu\nu\rho\sigma} A_a^{\nu} \left(\partial^\rho A_a^{\sigma} + 
\frac{1}{3} g f_{abc} A_b^{\rho} A_c^{\sigma}\right).
\label{KMU}
\end{equation}
Consequently it is suggestive to use $n\cdot K$ as a gluon polarization
partonometer due to its local character.  Corresponding calculations
in the framework of the bag model have been performed recently in
\cite{Jaffe96}.  Nevertheless, as we shall discuss, although
formally the operator $n\cdot K$ is built from local fields, its
matrix element is sensitive to large longitudinal distances in the same
 way as the matrix element  of the non-local operator in (\ref{ONL}).  
Thus the advantage of using $n\cdot K$ instead of 
$O_{\rm NL}$ is illusory.  
This is illustrated below within the framework of QCD sum rules.
We find that our estimates for
$\Delta G(\mu^2)$ as obtained from the gluon polarization
partonometer presented in \cite{MPS97}, and from 
the local operator (\ref{KMU}) differ
approximately by a factor of four.  
This discrepancy emphasizes, as we will show, the role
of contributions from different longitudinal length scales.

The remainder of this paper is organized as follows: in Sec.~2 we collect
the most important facts about the operator definition of the polarized gluon
distribution and the gluon polarization integral.  
In Sec.~3 we present a QCD sum rule  estimate 
for $\Delta G(\mu^2)$ starting out from the matrix element of 
$n \cdot K$ in light-cone gauge,
and review an estimate for $\Delta G(\mu^2)$ 
using the gluon partonometer introduced in Ref.\cite{MPS97}.  
We then discuss and explain in Sec.~4 the reasons for
the surprising discrepancy between the results obtained with these two methods.
Finally  Sec.~5 is devoted to a summary and conclusions.

\section{{Polarized gluon distribution  in QCD}}

In QCD parton distributions can be  related to matrix elements of twist-$2$  
non-local operators \cite{Collins}. 
In this framework unpolarized and polarized gluon distributions 
are defined through matrix elements of the light-cone string operators:
\begin{eqnarray} 
\label{eq:O(Delta,0)_unpol}
O_{G}(\Delta;0)& = & n_\mu n_\nu
{\mbox Tr}\, G^{\mu\xi}(\Delta)[\Delta;0]
{G}_\xi\,^\nu(0),
\\
\label{eq:O(Delta,0)}
O_{\Delta G}(\Delta;0) & = & n_\mu n_\nu
{\mbox Tr}\, G^{\mu\xi}(\Delta)[\Delta;0]
{\tilde G}_\xi\,^\nu(0).
\end{eqnarray}
Here $\Delta$ stands for a light-like vector being proportional to $n$.  
The path-ordered exponential  
\begin{equation}
[\Delta;0] = {\mbox P} \exp\left[ i g \Delta_{\mu} 
\int_0^1 d\lambda A^{\mu}(\Delta \lambda) \right],
\end{equation}
with the strong coupling constant $g$ and the gluon field 
$A^{\mu}$ guarantees gauge invariance  of the parton distributions 
(\ref{SMR1_unpol},\ref{SMR1}). 
The forward matrix elements of the string operators 
(\ref{eq:O(Delta,0)_unpol},\ref{eq:O(Delta,0)}) 
between nucleon states with momentum $p$ and spin $s$ 
define  the  unpolarized and polarized gluon distribution of a nucleon, 
$G(u,\mu^2)$ and $\Delta G(u,\mu^2)$,  
as a function of the Bjorken variable $u$\, and the normalization
scale $\mu^2$:
\begin{eqnarray}
\frac{1}{2} \sum_s\langle p,s | O_G(\Delta,0) | p,s \rangle_{\mu^2} 
 &=&  
(p\cdot n)^2 \int_0^1 \, du \, u \, G(u,\mu^2) 
\cos\left[u (p\cdot \Delta)\right],
\label{SMR1_unpol}
\\
\langle p,s | O_{\Delta G} (\Delta,0) | p,s \rangle_{\mu^2}  &=&  
(p\cdot n) (s \cdot n) \int_0^1 \, du \, u \, \Delta G(u,\mu^2) 
\sin\left[u (p\cdot \Delta)\right].
\label{SMR1}
\end{eqnarray}
The invariant measure of the light-cone distance between the 
two gluon fields in (\ref{SMR1_unpol},\ref{SMR1}) is given by the 
so called  Ioffe-time  $z = p \cdot \Delta$. 
Furthermore note that for a target polarized in the $3$-direction 
one has $s \cdot n = p \cdot n$.
Taking the Fourier transform of Eqs.~(\ref{SMR1_unpol}, \ref{SMR1}) 
yields the distribution functions: 
\begin{eqnarray}\label{eq:G}
u \,G(u,\mu^2) &=& \frac{1}{\pi \,(p\cdot n)^2} 
\int_0^{\infty} dz\,
\sum_s
\langle p,s | O_G(\Delta,0)  | p,s \rangle_{\mu^2} 
\cos(u z), 
\\ 
\label{eq:DG}
 u \,\Delta G(u,\mu^2) &=& \frac{2}{\pi\,(p\cdot n) (s\cdot n)} 
\int_0^{\infty} dz\,
\langle p,s | O_{\Delta G}(\Delta,0)  | p,s \rangle_{\mu^2} 
\sin(u z). 
\end{eqnarray}

These  definitions 
are of course in agreement with the perceptions of  the 
parton model.
Indeed in light-cone gauge, $n\cdot A = 0$, one can express 
the  distributions (\ref{eq:G},\ref{eq:DG}) in terms of  right- and 
left-handed gluon operators defined as 
$G^{+R(L)} = \varepsilon_{\mu}^{R(L)} G^{+ \mu}$,  
with the polarization vectors 
$\varepsilon^{\mu}_{R} = (0,-1,-i,0)/\sqrt{2}$ and 
$\varepsilon^{\mu}_{L} = (0,1,-i,0)/\sqrt{2}$:
\begin{eqnarray} 
u\,G(u,\mu^2) &=& \frac{1}{\pi\,p\cdot n} \int_0^{\infty} d \lambda\,
\cos{(p\cdot n \, \lambda u)} \nonumber \\
&\times& \sum_{s}
Tr \langle p,s | \left(G^{+ R} (n \lambda)\right)^\dagger  G^{+R} (0) +
\left(G^{+ L} (n \lambda)\right)^\dagger  G^{+L} (0) | p,s \rangle_{\mu^2},  
\nonumber \\
\\
u\,\Delta G(u,\mu^2) &=& \frac{2i}{\pi\,s\cdot n} \int_0^{\infty} 
d\lambda\, \sin{(p\cdot n \, \lambda u)} \nonumber \\
&\times & 
Tr \langle p,s | \left(G^{+ R} (n \lambda)\right)^\dagger  G^{+R} (0) -
\left(G^{+ L} (n\lambda)\right)^\dagger  G^{+ L} (0) | p,s
\rangle_{\mu^2}. \nonumber \\
\end{eqnarray}
After rewriting the gluon field strength tensor in terms of 
light-cone quantized fields one obtains:
\begin{eqnarray}
u\, G(u,\mu^2) &=& \int dx  \,d^2 k_{\perp} \delta (u-x) 
x\left[n_g (x, \vec k_{\perp},R,\mu^2) + 
n_g (x, \vec k_{\perp},L,\mu^2) \right],
\\
u\,\Delta G(u,\mu^2) &=& \int dx \,d^2 k_{\perp} \delta (u-x) 
x\left[ n_g (x, \vec k_{\perp},R,\mu^2) - 
n_g (x, \vec k_{\perp},L,\mu^2)\right], 
\end{eqnarray}
where $n_g(x,\vec k_{\perp},R(L),\mu^2)$ denotes the 
light-cone distribution 
function of right- (left-) handed gluons with light-cone momentum fraction 
$x$ and transverse momentum $\vec k_{\perp}$ 
\cite{Man90}. 

In the following we focus on the polarized gluon distribution.  
Performing a Taylor expansion of (\ref{SMR1}) around $\Delta = 0$ 
leads to well-known relations, or sum rules, between the moments of 
the polarized gluon distribution $\Delta G(u,\mu^2)$ and nucleon matrix 
elements of local QCD operators \cite{Alt81}:
\begin{eqnarray}
&&
\int_0^1 \, du \, u^{l-1} \, \Delta G(u,\mu^2)  
\equiv  
\Gamma_l(\mu^2), 
\quad \mbox{with} \,~l = 3,5,\dots,   
\nonumber \\ 
&&
n_\mu n_\nu {\mbox Tr}\, 
\langle p,s| G^{\mu \xi}(0) (in\cdot D)^{l-2} 
{\tilde G}_{\xi}\,^{\nu}(0) |p,s\rangle  
 =  (s\cdot n) (p\cdot n)^{l}\,
\Gamma_l(\mu^2).
\label{twist2} 
\label{SMR2}
\end{eqnarray}
Note, however, that a sum rule for $l=1$ 
which correspond to the integrated 
gluon polarization $\Delta G(\mu^2)$ is lacking. 
This is due to the fact  that a suitable gauge invariant, charge 
conjugation even, local operator which may serve as a 
gluon polarization partonometer does not exist. 

On the other hand we find from (\ref{SMR1}) that the 
polarized gluon distribution  $\Delta G(\mu^2)$ 
is determined by an integral over the corresponding 
Ioffe-time distribution which is defined as: 
\begin{equation}
\Gamma(z,\mu^2)  =  \int_0^1 du \,u \Delta G(u,\mu^2) \sin (u z). 
\label{ITD}
\end{equation}
Since in the distribution sense one has:   
\begin{equation} \label{principal_value}
\int_0^\infty dz \sin (u z) = 
\frac{1}{2} \left( \frac{1}{u + i \epsilon} + 
\frac{1}{u - i \epsilon}\right) = PV\frac{1}{u},
\end{equation}
where $PV$ denotes the principal value prescription, 
we indeed obtain:  
\begin{equation} 
\Delta G(\mu^2) = \int_0^\infty dz \, \Gamma(z, \mu^2) \, .
\label{DG}
\end{equation}
It is important to realize that 
$\Gamma(z,\mu^2) \sim z^{\alpha-2}$ for  large $z$, if 
$\Delta G(u,\mu^2)\sim u^{-\alpha}$ at small $u$. 
Therefore as long as $\alpha < 1$ the integral over the polarized 
gluon distribution (\ref{DG}) exists as it converges at large $z$ 
in an absolute sense. 
(At small $z$ the integrand in (\ref{DG}) should not cause harm since 
there $\Gamma(z) \approx z \,\Gamma_3$, 
with an anticipated finite third moment $\Gamma_3$.) 
Expanding the RHS of (\ref{ITD}) around $z=0$ yields a Taylor 
expansion of the Ioffe distribution $\Gamma(z,\mu^2)$ with 
coefficients proportional to the odd moments 
$\Gamma_l(\mu^2)$: 
\begin{equation} \label{eq:Gamma_expand}
\Gamma(z,\mu^2) = \Gamma_3 (\mu^2) \,z - 
\frac{1}{6} \Gamma_5(\mu^2) \,z^3
+ \frac{1}{120} \Gamma_7 (\mu^2) \,z^5  -  \dots
\end{equation}
Since each of these moments can 
be calculated, at least formally, as a reduced matrix element of a local, 
gauge invariant operator, the convergent integral (\ref{DG}) 
and hence $\Delta G(\mu^2)$ itself is a gauge-invariant quantity.

As already mentioned in the introduction, in light-cone gauge
the gluon polarization $\Delta G(\mu^2)$ can also be related 
to the expectation value of the topological current (\ref{KMU}): 
\begin{equation}
\langle p, s | n \cdot K(0) | p, s \rangle_{\mu^2} = 
(s \cdot n) \frac{\alpha_s}{2 \pi} \int_0^\infty dz \, \Gamma(z,\mu^2) = 
(s \cdot n) \frac{\alpha_s}{2 \pi} \Delta G(\mu^2).
\label{KMU1}
\end{equation}
Indeed, in light-cone gauge the 
operator $O_{\Delta G}(\Delta;0)$ in (\ref{eq:O(Delta,0)}) 
reduces to:
\begin{equation}
O_{\Delta G}(\Delta;0) = \partial^+ A^2(\Delta) \partial^+ A^1(0) - 
\partial^+ A^1(\Delta) \partial^+ A^2(0) \, ,
\end{equation}
where $A^1$ and $A^2$ denote the transverse components of the gluon field. 
This simplification occurs because in this gauge the path-ordered exponential
$[\Delta;0] = 1$, and the components
of the gluon field strength tensor which enter $O_{\Delta G}(\Delta;0)$ 
are given by $G^{+\perp} = \partial^+ A^\perp$. 
Using the definition (\ref{DG}) we obtain:
\begin{eqnarray}
\Delta G(\mu^2) &=&\frac{1}{2 (s\cdot n)(p \cdot n)} \int_0^\infty 
dz \,\langle p,s|\partial^+ A^2(\Delta) \partial^+ A^1(0) - 
\partial^+ A^1(\Delta) \partial^+ A^2(0) |p,s\rangle_{\mu^2}
\, , \nonumber \\
&=&
\frac{1}{2 (s\cdot n)} \int_0^\infty d\lambda \, 
\langle p,s|
\frac{\partial}{\partial \lambda} A^2(\lambda n) \partial^+ A^1(0) - 
\frac{\partial}{\partial \lambda}A^1(\lambda n) \partial^+ A^2(0) 
|p,s\rangle_{\mu^2}.
\label{KMU2}
\end{eqnarray}
Integrating by parts and assuming that the
boundary term vanishes for $\lambda \to \infty$, one finally obtains:
\begin{equation}
(s \cdot n) \Delta G(\mu^2) =
\langle p,s| A^1(0) \partial^+ A^2(0) - A^2(0) \partial^+ A^1(0) |p,s\rangle, 
\end{equation}
which is easily shown to be equivalent to (\ref{KMU1}). This relation
involves an operator which is built from local fields. 
Therefore its matrix element is calculable using known methods to deal with
local operators. This fact however turns out to be of no real
advantage as we will point out in Sec.~4, where we 
describe a QCD sum rule calculation.

To complete our discussion we derive the scale dependence of 
the gluon polarization  $\Delta G(\mu^2)$ starting out from 
the operator definition (\ref{DG}).
The one-loop evolution equation for
the Ioffe-time distribution $\Gamma(z,\mu^2)$ reads \cite{BGR97,BR97}:
\begin{equation}
\mu^2 \,\frac{d \Gamma(z,\mu^2)}{d\mu^2} = 
\frac{\alpha_s(\mu^2)}{2 \pi}\int_0^1 dv [K_1(v) \Gamma(vz,\mu^2) +
K_2(v) \frac{1}{z} \Sigma(vz,\mu^2)],
\label{EV1}
\end{equation}
where: 
\begin{eqnarray}
K_1(v) & = & \frac{\beta_0}{2} \delta({\bar v}) + 2 C_A 
\left(2 v {\bar v} +
\frac{1}{{\bar v}_+} - 1 - v\right), \nonumber \\
K_2(v) & = & - C_F \left(\delta({\bar v}) - 2 {\bar v}\right),
\label{K12}
\end{eqnarray}
with $C_F = {4}/{3}$, $C_A = 3$, $\beta_0 = 11 - \frac{2}{3} N_f$, with $N_f$
being the number of active flavors, and ${\bar v} = 1- v$. The factor ${1}/{z}$
in (\ref{EV1}) arises in accordance with the definition of $\Sigma(z,\mu^2)$,
the Ioffe-time distribution of polarized quarks:
\begin{eqnarray}
\langle p,s|{\bar \Psi}(\Delta) {\hat n} \gamma_5 
[\Delta;0]
\Psi(0)|p,s\rangle & + &
(\Delta \to - \Delta) = 4 (s \cdot n) \Sigma(z,\mu^2), \nonumber \\
\mbox{with}\quad
\Sigma(z,\mu^2) & = & \int_0^1 du \,\Delta q(u,\mu^2) \cos (u z) \, .
\label{SIG1}
\end{eqnarray}
Here $\Delta q(u,\mu^2)$ is the flavor-singlet polarized quark distribution
in momentum space.  
The evolution equation for $\Delta
G(\mu^2)$ is obtained by integrating both sides of Eq.~(\ref{EV1}) 
over $z$. 
We find that gluons enter only via the term in $K_1$ being proportional 
to $\beta_0$.
Furthermore the 
quark contribution can be transformed  conveniently using the identity:
\begin{equation}
\frac{1}{z} \int_0^1 dv \left(\delta ({\bar v}) - 2 {\bar v}\right) 
\cos{(u v z)} = - 2 u \int_0^1 dv v \left ( 
1- \frac{1}{2} v\right) \sin{(u v z)}. 
\end{equation}
Applying Eq.~(\ref{principal_value}) yields then 
the standard one-loop evolution equation \cite{Alt81}:
\begin{eqnarray}\label{EV2}
\frac{d \Delta G(\mu^2)}{d t} & = & \frac{\alpha_s}{2 \pi} 
\left( \frac{\beta_0}{2}
\Delta G(\mu^2) + \frac{3}{2} C_F \Sigma(\mu^2)\right), \\
\mbox{with}\quad t & = & \log(\mu^2), \;\mbox{and}\;\; 
\alpha_s = \frac{4\pi}{\beta_0 t}. \nonumber
\end{eqnarray}
Here 
\begin{equation}
\Sigma(\mu^2) = \int_0^1 du \,\Delta q(u,\mu^2)
\label{SIG2}
\end{equation}
is the quark polarization in the proton\footnote{For considerations of
the one-loop evolution equation for $\Delta G(\mu^2)$ it is not
necessary to discuss the role of the axial anomaly in the
interpretation of $\Sigma(\mu^2)$.}.  At the one loop level one has
\cite{Alt81}:
\begin{equation}
\frac{d \Sigma(\mu^2)}{d t}  =  0, \quad\mbox{and thus} \;\;\;\;
\Sigma(\mu^2)  = \Sigma_0 = {\rm constant},
\end{equation}
i.e. to this accuracy the quark polarization is scale-independent.
One then obtains as a solution of (\ref{EV2}) the well known 
result \cite{Alt81,Ratc89}:
\begin{equation}
\Delta G(\mu^{\prime 2}) = 
\frac{\alpha_s(\mu^2)}{\alpha_s(\mu^{\prime 2})} 
\,\Delta G(\mu^2)
+ \frac{4}{\beta_0} 
\Sigma_0 \left(\frac{\alpha_s(\mu^2)}{\alpha_s(\mu^{\prime 2)}} -1
\right).
\end{equation}
Thus we have shown that the operator definition of $\Delta G(\mu^2)$ 
is indeed equivalent  to parton model considerations.

\section{{\bf Different gluon polarization partonometers}}

As discussed above, the gluon polarization 
in the nucleon can be obtained either from the integral 
over the Ioffe-time distribution (\ref{DG}), or 
in light-cone gauge 
from the forward matrix element of the topological current 
(\ref{KMU1}). 
Here we explore and compare both approaches in the 
framework of QCD sum rules.

\subsection{Gluon polarization from the topological current}
\label{ssec:Gpol_nK}

To estimate the  nucleon matrix element of the topological 
current (\ref{KMU1})  
we perform a standard QCD sum rule calculation. 
For this purpose we consider the three-point correlation function
\begin{equation}
I_{K}= i^2\!\int d^4 x\, e^{iq\cdot x}\int d^4y \, e^{ip\cdot y}
\langle 0| T[ \eta_G(x) \, n\cdot K(y) \,\bar{\eta}_G(0) ]
|0\rangle  
\label{kmucor}
\end{equation}
of the operator $n \cdot K$ in light-cone gauge and the nucleon interpolating 
currents $\eta_G$, ${\bar \eta}_G$. For the latter we take:
\begin{eqnarray}
\eta_G(x) & = & \frac{2}{3} (\eta_{\rm old}(x) - \eta_{\rm ex}(x)), 
\\
\eta_{\rm old}(x) & = & \epsilon^{abc} (u^{a T}(x)C\gamma_\mu u^b(x)) \gamma_5
\gamma^\mu \sigma_{\alpha\beta}
\left[ {g} G^{\alpha\beta}(x) d(x) \right]^c,
\nonumber \\
\eta_{\rm ex}(x) & = & \epsilon^{abc} (u^{a T}(x)C\gamma_\mu d^b(x)) \gamma_5
\gamma^\mu \sigma_{\alpha\beta}\left[ {g} 
G^{\alpha\beta}(x) u(x) \right]^c. 
\nonumber
\label{etaG}
\end{eqnarray}
Its overlap with the nucleon state,
\begin{equation}
\langle 0|\eta_G(0)|p,s\rangle = m_N^2 \lambda_G u(p,s) \, ,
\label{overlap}
\end{equation} 
at the scale $\mu^2 \sim 1$ GeV$^2$ has been determined in 
Ref.\cite{BGMS93}.
Note that this current, which contains explicit gluon degrees of 
freedom, has been successfully employed in investigations of 
nucleon matrix elements of QCD operators being 
sensitive to  gluon components of the nucleon wave
function \cite{BGMS93,Ste95,Bel96}. 
To avoid large t-channel contributions we 
stay at Euclidean momenta 
$Q^2 = - q^2 \approx (1 - 4)$ GeV$^2$ and perform a 
numerical extrapolation to $Q^2 = 0$ at the end. 
Furthermore the kinematic is chosen  such that 
$q\cdot n = 0$, i.e. $q^2 = - \vec q_\perp\,^2$.

In the following we concentrate on the contribution to 
the correlator (\ref{kmucor}) of the form:\footnote{
We suppress here and in the following any  dependence on the 
scale $\mu^2$.}
\begin{equation}
\gamma_5 \,{\hat p} \,(p\cdot n) T_{K}(p^2,(p+q)^2,Q^2)\, .
\end{equation}
The invariant function $T_{K}$ can be projected out uniquely 
from $I_{K}$. It receives contributions from nucleons as well 
as from higher resonances and continuum states. 
For the nucleon contribution we have:
\begin{eqnarray}
T_{K}(p^2,(p+q)^2,Q^2) 
 = 
\frac{\mbox{Tr} \left( \hat n \gamma_5 I_{K}\right)} 
{ 4 (p\cdot n)^2} 
=
\frac{\lambda_G^2 \,m_N^4}{(m_N^2 - p^2)(m_N^2 - (p+q)^2)} 
\frac{\alpha_s}{2\pi}\Delta {\tilde G}(Q^2),
\label{nucleon}
\end{eqnarray}
where the form factor $\Delta {\tilde G}(Q^2)$  
coincides at $Q^2=0$ with the gluon polarization $\Delta G$.

As a next step we use the fact that 
$T_{K}$ admits a double spectral representation 
\cite{many}:
\begin{equation}
T_{K}(p^2,(p+q)^2,Q^2) = 
\int \frac{ds_1}{s_1-p^2}
\int \frac{ds_2}{s_2-(p+q)^2} \;\rho_{K}(s_1,s_2,Q^2).
\label{disp}
\end{equation}
We have calculated the spectral density $\rho_{K}$ in light-cone 
gauge, taking into account the dimension-$1$ and dimension-$6$ 
operators in the OPE of $T_{K}$, which are visualized in Fig.~1. 
We obtain:
\begin{eqnarray}
\rho^{(1)}_K(s_1, s_2, Q^2) &=& 
\frac{\alpha_s\,Q^2\,\left( \Delta_q - R^{1\over 2} \right)^2\,}
 {23040\,{\pi}^5\,R^{5\over 2}}
      \left( 
  \Delta_q\, Q^2 \, R^{3\over 2} - 
  3\, \Delta_q\, R^2  - 
  Q^2\, R^2 
\right.
\nonumber \\
&-&
\left.
  7\, R^{5\over 2} - 
  2\, \Delta_q\, Q^2 \, R^{1\over 2}\, s_1\, s_2 + 
  2\, \Delta_q\, R\, s_1\, s_2 + 
  4\,R^{3\over 2}\, s_1\, s_2 - 
  4\, Q^2\, s_1^2 \, s_2^2
      \right), 
\nonumber \\
\rho^{(6)}_K(s_1, s_2, Q^2)&=& 
\frac{56\,\alpha_s}{9\,\pi}
\, {\langle {\bar q} q \rangle}^2
\,\frac{\Delta_q\,Q^4\,s_1\,s_2}{ R^{5\over 2}},
\label{rho_K_def}
\end{eqnarray}
where 
$\Delta_q = s_1 + s_2 + Q^2$
and
$R = \Delta_q^2 - 4 s_1 s_2$.
To eliminate contributions from higher resonances 
and the continuum we limit the integral over $s_1$ and $s_2$ by 
the continuum threshold $s_0$. 
Performing in addition a Borel transformation  
in $p^2$ and $(p+q)^2$ yields the sum rule:
\begin{equation}  \label{sum_rule}
\frac{\alpha_s}{2 \pi} \Delta {\tilde G}(Q^2) = 
\frac{e^{m_N^2/M^2}}{\lambda_G^2 m_N^4} 
\int_0^{s_0} d s_1 \int_0^{s_0} d s_2 \,e^{-(s_1+s_2)/2M^2} 
\rho_{K}(s_1,s_2,Q^2) \, ,
\label{sum_rule_K}
\end{equation}
where $M^2$ is the Borel parameter. 
For consistency both, $s_0$ and $M^2$ should be taken around their 
values fixed by the two-point sum rules \cite{BGMS93}. 
In the actual calculation we use at the scale  $\mu^2 = 1\,{\rm GeV}^2$
the standard value  for the quark condensate,  
$-(2\pi)^2\langle \bar q q\rangle = 0.67\,{\rm GeV}^3$, 
and the strong coupling  $\alpha_s = 0.37$ \cite{BGMS93}.

The stability of the sum rule (\ref{sum_rule}) against variations of the Borel
parameter $M^2$ and the continuum threshold $s_0$ is illustrated in Figs.~2 and
3. One observes a relatively strong dependence on $s_0$ which can be 
traced back
to the large dimension of the interpolating current $\eta_G$.  To determine
the gluon polarization $\Delta G$ we extrapolate the form factor $\Delta
{\tilde G}(Q^2)$ to $Q^2 \rightarrow 0$.  For this purpose we fit the RHS of
(\ref{sum_rule}) in the interval $1 \, {\rm GeV}^2 \le Q^2 < 4 \, {\rm
  GeV}^2$ by:
\begin{equation} 
\Delta {\tilde G}(Q^2) =  
{\Delta G} \,\frac{1}{[1+Q^2/m^2]^\gamma}\, ,
\label{fit}
\end{equation}
with $\gamma = 3$ as suggested by quark counting rules.  
Extrapolating to $Q^2 \to 0$ gives $\Delta G(\mu^2 \sim 1\, {\rm GeV}^2)
= 0.6 \pm 0.2$, where the error has been estimated from the $M^2$ and $s_0$
dependence of the extrapolation. We have checked that this result essentially
does not change if we allow $\gamma$ as a fit parameter as well. 
An additional
$30 \%$ error arises from the uncertainty in $\lambda_G$ and the 
vacuum saturation ansatz for the four-quark condensate \cite{BGMS93}.

\subsection{Gluon polarization from an approximate Ioffe-time distribution} 
\label{ssec:Gpol_Ioffe}

In comparison to the above calculation we review here an estimate 
of $\Delta G$ 
via an approximate Ioffe-time distribution \cite{MPS97}.  
The latter 
is based on the conjecture that in the laboratory frame the polarized
gluon distribution (\ref{DG}) receives major  contributions only from
longitudinal distances smaller than the nucleon diameter, as
determined by the nucleon electromagnetic form factor.  
This  hypothesis is supported both by Regge
phenomenology \cite{Regge} and the color coherence hypothesis \cite{Bro95},
which impose strong restrictions on polarized glue at small $u$ or,
equivalently, at large $z$.  
For a more accurate determination of 
the longitudinal length scale at which the 
gluon polarization (\ref{DG}) effectively saturates 
an additional  assumption has to be made.
In this respect we have constructed in 
\cite{MPS97} a polarized gluon correlation
function $\Gamma(z)$ which behaves at large $z$ 
similar to  the valence quark distribution  of a  nucleon, 
i.e. it becomes small for $z\gsim 10$. 
(In the nucleon rest frame $z=10$ corresponds 
to a longitudinal distance of $2\,{\rm fm}$.)  
Indeed most of the current parametrizations for polarized glue sustain
this picture \cite{Lad96}, as shown in Fig.~4 for the distributions 
from Refs.\cite{Bro95,Chiap}.

In Fig.~5 we present the Ioffe-time
distribution as obtained from the expansion in Eq.~(\ref{eq:Gamma_expand}),
taking into account terms up to the $l$-th moment of $\Delta G(u)$.  
One finds that the first two non-vanishing moments 
$\Gamma_3$ and $\Gamma_5$ determine the Ioffe-time distribution 
$\Gamma(z)$ at small $z$ nearly up to its maximum\footnote{Note 
that in the notation of \cite{MPS97} $\Gamma_3$ and
  $\Gamma_5$ correspond to $\Gamma_2$ and $\Gamma_4$, respectively.}.
Therefore if the polarized gluon distribution is of regular shape
similar to the ones shown in Fig.~4, and obtains
significant contributions only from the region $z \lsim 10$, a simple
estimate of $\Delta G(\mu^2)$ is possible \cite{MPS97}.  
It is given by the area of the
triangle spanned by the points $z=0$, $z=10$, and the maximum of the
approximate Ioffe-time distribution
\begin{equation} \label{eq:Ioffe_approx}
\Gamma(z) \approx \Gamma_3 \,z - 
\frac{1}{6} \,\Gamma_5 \,z^3.
\end{equation}
This rough estimate requires only 
the knowledge of two moments of 
$\Delta G(u)$.
These can, as a matter of principle, be taken from any 
theoretical investigation, 
i.e. QCD sum rules or lattice calculations. 

Of course one can do better  
if the normalization of $\Gamma(z)$ is known at 
some large value of $z\approx 10$. 
Then a more accurate estimate 
can be achieved, calculating the area bound by 
the approximate Ioffe-time distribution (\ref{eq:Ioffe_approx}) 
up to its maximum  and a straight line connecting this point with 
the value of $\Gamma(z=10)$.
For the model parametrizations \cite{Bro95,Chiap} this leads 
to an estimate for $\Delta G$ with $(10 - 20)\%$ accuracy.
In \cite{MPS97}  the color coherence hypothesis 
\cite{Bro95} has been applied to obtain information on the 
large $z$ or, equivalently,  small $u$ behavior of polarized 
glue in nucleons:
\begin{equation}
\frac{\Delta G(u)}{G(u)}
\rightarrow {u}, 
\quad \mbox{for} \;u\to 0.
\label{coh}
\end{equation}
It allows to estimate $\Gamma(z=10)$ from parametrizations 
for the  unpolarized gluon distribution. For example one finds 
from the GRV \cite{GRV} and CTEQ \cite{CTEQ} 
LO unpolarized gluon distributions 
$\Gamma(z=10,\mu^2\sim 1\,{\rm GeV}^2) = 0.005 - 0.007$.
On the other hand  the 
parametrizations of \cite{Bro95} and \cite{Chiap} yield
$\Gamma(z=10,\mu^2\sim 1\,{\rm GeV}^2) = 0.02$ and $0.09$,
respectively. Since in principle 
nothing prevents $\Gamma(z)$ from becoming
negative at large $z$, $ -0.05 \le \Gamma(z=10) \le 0.05$ should be a
conservative estimate.

The moments $\Gamma_3$ and $\Gamma_5$ have been calculated within  
a standard QCD sum rule approach, 
starting from  an investigation of the  three-point correlation 
function \cite{MPS97}:
\begin{eqnarray}
I_{\Gamma} & = & i^2\!\int d^4 x\, e^{iq\cdot x}\int d^4y \, e^{ip\cdot
y} \langle 0| T[ \eta_G(x) O_{\Delta G}(y+\Delta/2;y-\Delta/2) 
\bar{\eta}_G(0) ]
|0\rangle, \nonumber \\
& = &   
\gamma_5 \,{\hat p} \,(p\cdot n)^2 \,T_{\Gamma}(p^2,(p+q)^2,Q^2, z) 
+ \dots.
\label{thetacor}
\end{eqnarray}
Separating the contribution from  nucleon states yields:
\begin{eqnarray}
T_{\Gamma}(p^2,(p+q)^2,Q^2, z) =  
\frac{\mbox{Tr} \left( \hat n\gamma_5 I_{\Gamma}\right)}
{4 (p\cdot n)^3} 
=  
\frac{\lambda_G^2 \,m_N^4}{(m_N^2 - p^2)(m_N^2 - (p+q)^2)} 
{\tilde \Gamma}(z,Q^2) + \dots ,
\label{nucleon1b}
\end{eqnarray}
where the form factor ${\tilde \Gamma}(z,Q^2)$
coincides at $Q^2 = 0$ with the Ioffe-time distribution 
$\Gamma(z)$ in (\ref{ITD}).  
In a next step again a double spectral representation 
is introduced:
\begin{equation}
T_{\Gamma}(p^2,(p+q)^2,Q^2,z) = 
\int \frac{ds_1}{s_1-p^2}
\int \frac{ds_2}{s_2-(p+q)^2} \;\rho_{\Gamma}(s_1,s_2,Q^2,z).
\label{disp1b}
\end{equation}
As in Sec.~3.1 
the dimension-$1$ and dimension-$6$ contributions 
have been considered.
For them the polarized spectral density 
$\rho_{\Gamma}$ coincides with the
spectral density for the operator (\ref{eq:O(Delta,0)_unpol}) which 
determines the unpolarized gluon distribution $G(u,\mu^2)$ \cite{BGMS93}.  
Since this spectral density leads to a realistic large value 
for the contribution of gluons to the nucleon momentum, it is suggestive
that also the integrated gluon polarization is large.

Expanding Eqs.~(\ref{nucleon1b},\ref{disp1b}) 
in powers of $z$ yields sum rules for
$\Gamma_3(Q^2)$ and $\Gamma_5(Q^2)$, which are given  
explicitly in \cite{MPS97}.  
An extrapolation to $Q^2 = 0$ as in (\ref{fit})
finally leads to  the desired moments.  
In combination with the color coherence hypothesis (\ref{coh}) one
then obtains $\Delta G(\mu^2 \sim 1 \, \rm{GeV}^2) = 2\pm 1$.  The main
source of the quoted error is due to uncertainties in the QCD sum rule 
approach.  Furthermore, under the assumption that $\Gamma(z)$ 
follows the Regge
behavior at large $z$,  possible contributions from longitudinal distances
beyond the ones accounted for have been estimated to be smaller than 
$\pm 0.2$.
Note however that current experience \cite{BGM95,Ross96} indicates that QCD
sum rules significantly overestimate higher moments of parton
distribution functions.  Therefore the result quoted above should be treated 
as an upper limit for $\Delta G(\mu^2 \sim 1 \, \rm{GeV}^2)$.

\section{Discussion}

We have found that the gluon polarization obtained from the matrix element of
the topological current differs by a factor of around four as
compared to the estimate based on the approximate Ioffe-time distribution.
This obviously calls for an explanation.  As we show below, this discrepancy
emphasizes the role played by contributions from different longitudinal
distances.

First let us comment on the question whether we have gained anything by going
from the non-local definition of $\Delta G$ to the local one. 
To this end note  that due to Eq.~(\ref{KMU1}) one has:\footnote{
We remind the reader that this relation holds only if the gauge-dependent 
density $\rho_K$ is calculated in light-cone gauge.}
\begin{equation}
\rho_{K}(s_1,s_2,Q^2) = \int_0^\infty dz \, 
\rho_{\Gamma}(s_1,s_2,Q^2,z) \, . 
\label{relation}
\end{equation}
Of course this relation can be directly verified for 
the contributions  from the operators of dimension 
one  and six as used in  Sec.~3.1 and 3.2. 
This however implies that, in accordance with the equivalence
of the definitions (\ref{DG}) and (\ref{KMU1}), 
the matrix element of $n \cdot K$ is as sensitive to 
different longitudinal distances as the matrix element of the
non-local operator (\ref{eq:O(Delta,0)}).
Consequently the apparent local form of $K$ does
not remove the essential 
non-local character  of $\Delta G$.

To clarify the reason for our different results for $\Delta G$ 
consider a polarized gluon distribution with the following behavior:
\begin{equation}
G(u)  \sim
\left\{\begin{array}{l} 
{\cal A} \, u^{-\alpha}, \hspace*{0.76cm}\quad\mbox{for} \quad u \to 0, \\ 
\nonumber \\
{\cal B} \, (1-u)^\beta, \quad\mbox{for}  \quad u \to 1 \, .
\end{array}\right.
\end{equation}
The asymptotic form of the corresponding Ioffe-time distribution at large
$z$ arising from the regions $u \to 0$ and $u \to 1$ reads:
\begin{equation}
\Gamma(z)  =  {\cal A}
\sin{\left(\frac{\pi}{2} \alpha\right)} 
\Gamma_{\rm E}(2-\alpha) z^{\alpha-2} + \dots -
{\cal B} \cos{\left(z-\beta \frac{\pi}{2}\right)} 
\Gamma_{\rm E}(\beta+1) z^{-1-\beta} 
+ \dots,
\label{asym} 
\end{equation}
where $\Gamma_{\rm E}$ denotes the Euler Gamma function.
Phenomenologically one would expect $0 < \alpha < 1$ and $\beta \ge 4$ 
\cite{Bro95}, which guarantees a smooth non-oscillatory behavior of 
$\Gamma(z)$ at large distances. 
(In  (\ref{asym})
even for $\alpha \sim 0$ the next-to-leading contribution from the
small $u$ region dominates over the leading one arising from 
the region $u \to 1$, if only $\beta$ is large enough and 
${\cal A}$ and ${\cal B}$ are of similar magnitude.) 

The Ioffe-time distribution obtained from QCD sum rules behaves 
however differently.
This can be seen from Fig.~6 where we show its dominant contribution 
which results from the dimension-$6$ operator 
taken in the limit $Q^2 \rightarrow 0$.  
This limit should give a qualitatively reasonable estimate since the
$Q^2$-dependence of the sum rules in Sec.~\ref{ssec:Gpol_nK} and
\ref{ssec:Gpol_Ioffe} has turned out  to be quite smooth.
We find that, contrary to what one expects from
phenomenological considerations, $\Gamma (z)$ oscillates strongly and 
decreases relatively slow at large $z$. 
In terms of the characteristic exponents in
(\ref{asym}) its behavior corresponds approximately to $\alpha \sim - 1$
and $\beta \sim 0$.
This shows that QCD sum rules yield a 
distribution skewed towards large values of $u$, and consequently leads
to predictions for  the moments of $\Delta G(u)$ which are too large. 
The small value of $\beta$
reflects a large weight for configurations in which one gluon carries 
most of the momentum of the nucleon. 
In the present calculation 
this is due to the fact that 
neither a perturbative (Sudakov), nor a non-perturbative 
(large invariant mass) suppression mechanism 
for such configurations is present in lowest order OPE. 
Since the sum rule for $n \cdot K$ in
Sec.\ref{ssec:Gpol_nK} accounts for the full integral over $\Gamma (z)$ from
zero to infinity, it incorporates all oscillations at intermediate and large
$z$, leading to a small value of $\Delta G$.  
Apart from the small-$z$ region the $z$-dependence of
$\Gamma(z)$ shown in Fig.~6 is certainly not realistic.
Therefore  we have to conclude that the QCD sum rule  calculation of the 
matrix element of $n \cdot K$ is not entirely self-consistent, as it 
receives large contributions from regions
where the sum rule method is not reliable.

This is different in the approach discussed in Sec.~\ref{ssec:Gpol_Ioffe}. 
Here we have assumed that QCD sum rules yield a reasonable estimate for 
the first two moments $\Gamma_3$ and $\Gamma_5$.
However we have discarded oscillations of $\Gamma(z)$ 
at large $z$, assuming that the  main contribution to 
$\Delta G$ arises from small light-cone distances.
As a consequence QCD sum rules are used only in a domain where they are in 
principle applicable. 
We therefore believe that such an estimate of $\Delta G$ is 
better justified.

\section{Summary and conclusions}

In QCD the twist-2 polarized gluon distribution is defined 
through  a gauge
invariant but non-local string operator.  As a consequence $\Delta G$
can receive, at least in principle, contributions from different
longitudinal distances.  Although in light-cone gauge $\Delta G$ can
be formally expressed through the forward matrix element of the 
local topological current, also in this
case contributions from all longitudinal distances are accumulated.
Therefore the latter can be used for a trustworthy estimate of $\Delta
G$ only if an approximation to strong interaction dynamics 
is available which is
applicable at all longitudinal distances.  We have
illustrated this point in the framework of a QCD sum rule  calculation 
which leads to rather unrealistic  contributions from large
longitudinal distances, and results in a small value for $\Delta
G(\mu^2 \sim 1\,{\rm GeV}^2) = 0.6 \pm 0.2$.

For a self consistent estimate of the gluon polarization in nucleons 
one has to ensure that the main support to $\Delta G$ results from distances 
where the used approximation is supposed to do its best. 
In the case of QCD sum rules only contributions from small distances 
can be approximated in a reasonable way. 
Combining the latter with the assumption that contributions to $\Delta G$ 
from distances larger than the typical nucleon size are small yields 
$\Delta G(\mu^2 \sim 1\, {\rm GeV}^2) = 2\pm 1$.

If, contrary to our assumption, $\Delta G$ receives 
at low normalization scales $\mu^2$ 
important contributions from large longitudinal distances, 
new interesting questions would arise 
since such contributions could
hardly be interpreted as being due to confining glue, 
understood as part of a nucleon with a size of around one fm. 
At large normalization scales GLAP evolution may result in a more
and more singular behavior of $\Delta G(u,\mu^2)$ at small values of $u$. In
this case it is understood, however, that such contributions have to be 
filtered out in order to learn something about the distribution 
of nucleon polarization among low-virtuality degrees of freedom.

To avoid in the calculation of moments of a parton distribution 
a situation in which an approximation tailored for  contributions 
from small 
longitudinal distances generates large contributions from distances beyond 
its scope of applicability, one should require that it gives
a reasonable behavior of the considered distribution at large
values of $z$ or, equivalently, at values of $u$ close to $0$\footnote{To 
   achieve this for example 
   for the quark distribution in the bag model 
   one has to introduce a complicated projection which allows to
  approximately express bag model solutions in terms of eigenfunctions of
  the momentum operator \cite{Tony95}.}.
If such a requirement cannot be fulfilled
an explicit construction  of  contributions from large longitudinal 
distances , like in \cite{MPS97}, is necessary.

\bigskip
\bigskip

{\bf Acknowledgments}: 
We gratefully acknowledge discussions with V. Braun about
calculations presented in this paper. This work was supported in part by BMBF,
KBN grant 2~P03B~065~10, and German-Polish exchange program X081.91.

\vfill
\eject

\newpage



\begin{minipage}{16cm}
\vspace{5cm}
\bild{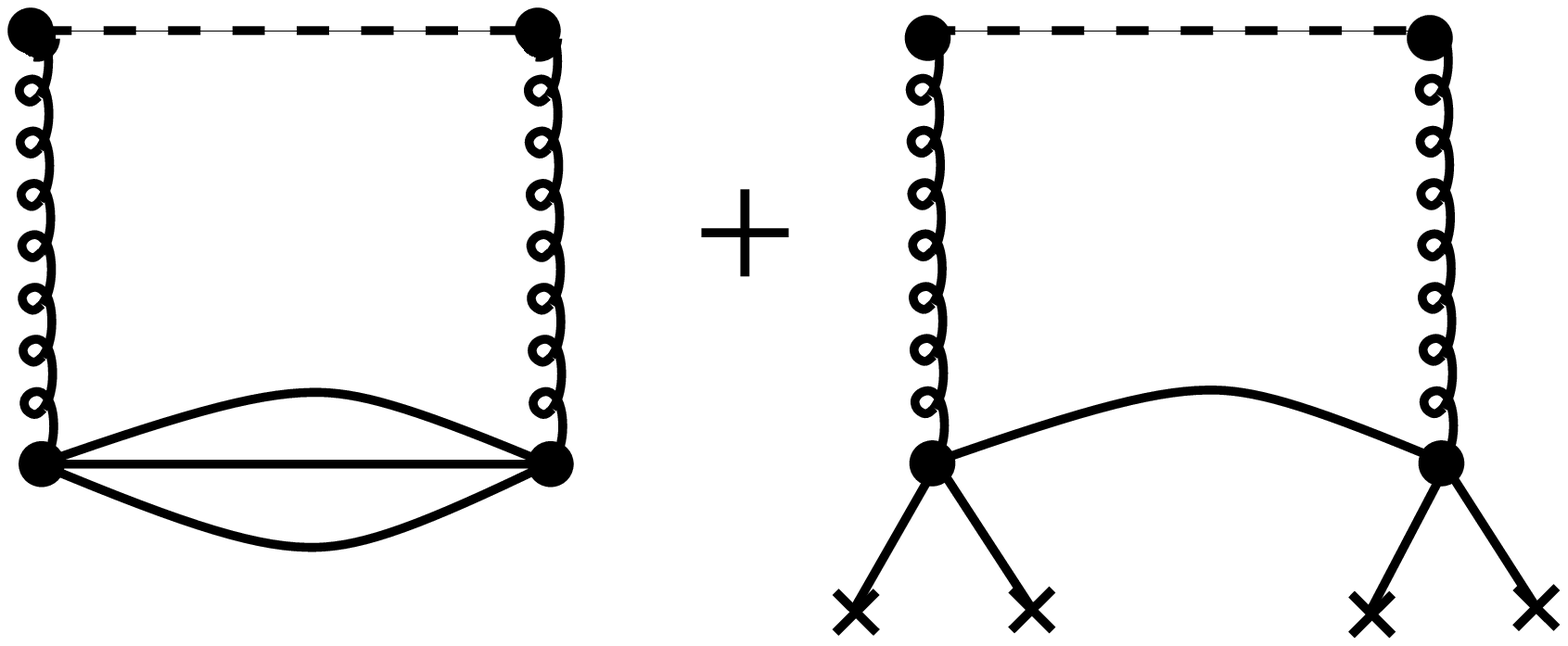}{13}
\label{fig_graphs}
\end{minipage}
\begin{description}
  
\item[Fig.~1:] 
Graphs representing the contributions from dimension-$1$ and 
dimension-$6$ operators 
to the QCD sum rule calculations described  in the text.  
\end{description}
\pagebreak


\begin{minipage}{16cm}
\bild{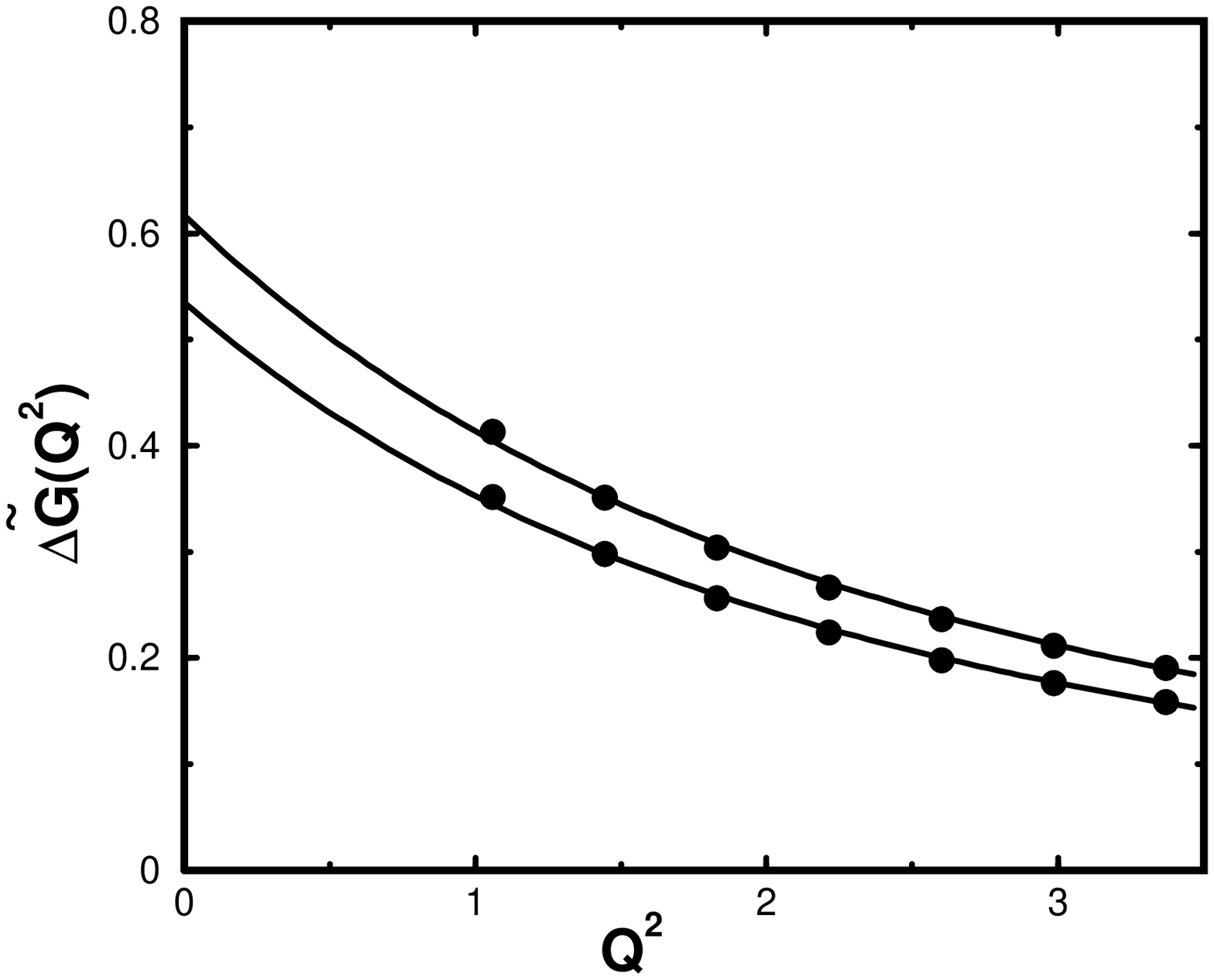}{13}
\label{fig_stab_M2} 
\end{minipage}
\begin{description}
  
\item[Fig.~2:] 
Stability of the sum rule for the nucleon matrix element of 
   $n\cdot K$ against variations of the Borel mass between  
   $M^2=1\, {\rm GeV}^2$ (lower curve) 
   and $M^2=2\, {\rm GeV}^2$ (upper curve). 
   The continuum threshold has been fixed at ${\sqrt s_0} = 1.5 \, 
   {\rm GeV}$.  
   The dots  correspond to the
   sum rule results (\ref{sum_rule_K}) for  $ 1\, {\rm GeV}^2 \le Q^2 < 4 \,
  {\rm GeV}^2$. 
   The solid lines show  the extrapolation to 
   $Q^2 \rightarrow 0$ (\ref{fit}).
\end{description}
\pagebreak

\begin{minipage}{16cm}
\bild{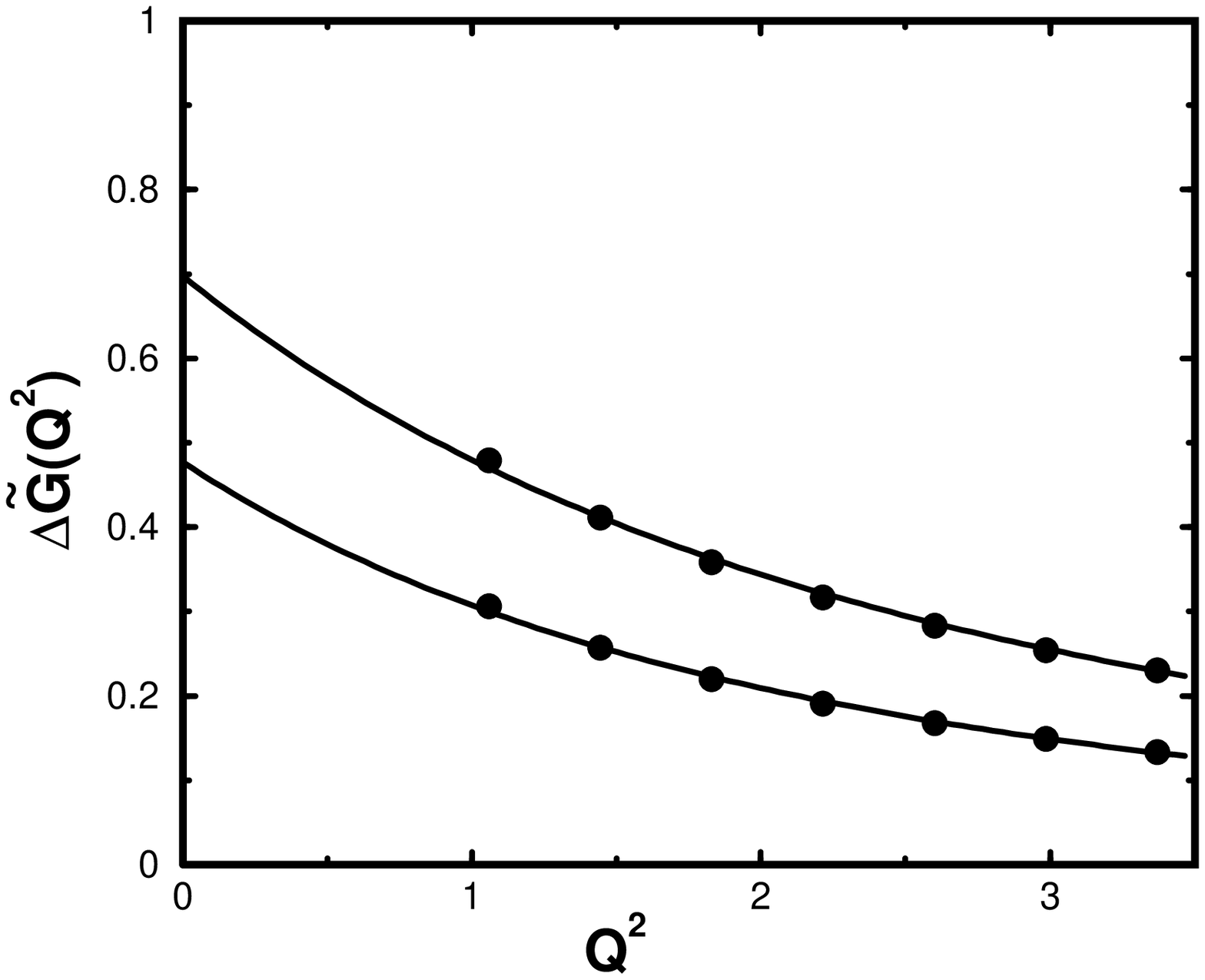}{13}
\label{fig_stab_s_0}
\end{minipage}
\begin{description}
  
\item[Fig.~3:] 
Stability of the sum rule for the nucleon matrix element of 
   $n\cdot K$ against variations of  the continuum threshold between 
   $\sqrt s_0=1.4\,
   {\rm GeV}$ (lower curve) and $\sqrt s_0=1.6\, {\rm GeV}$ (upper curve). 
   The Borel
   mass has been fixed at $M^2 = 1.5$ GeV$^2$. 
   The dots  correspond to the
   sum rule results (\ref{sum_rule_K}) for  $ 1\, {\rm GeV}^2 \le Q^2 < 4 \,
   {\rm GeV}^2$. 
   The solid lines show  the extrapolation to 
   $Q^2 \rightarrow 0$ (\ref{fit}).
\end{description}
\pagebreak


\begin{minipage}{16cm}
\bild{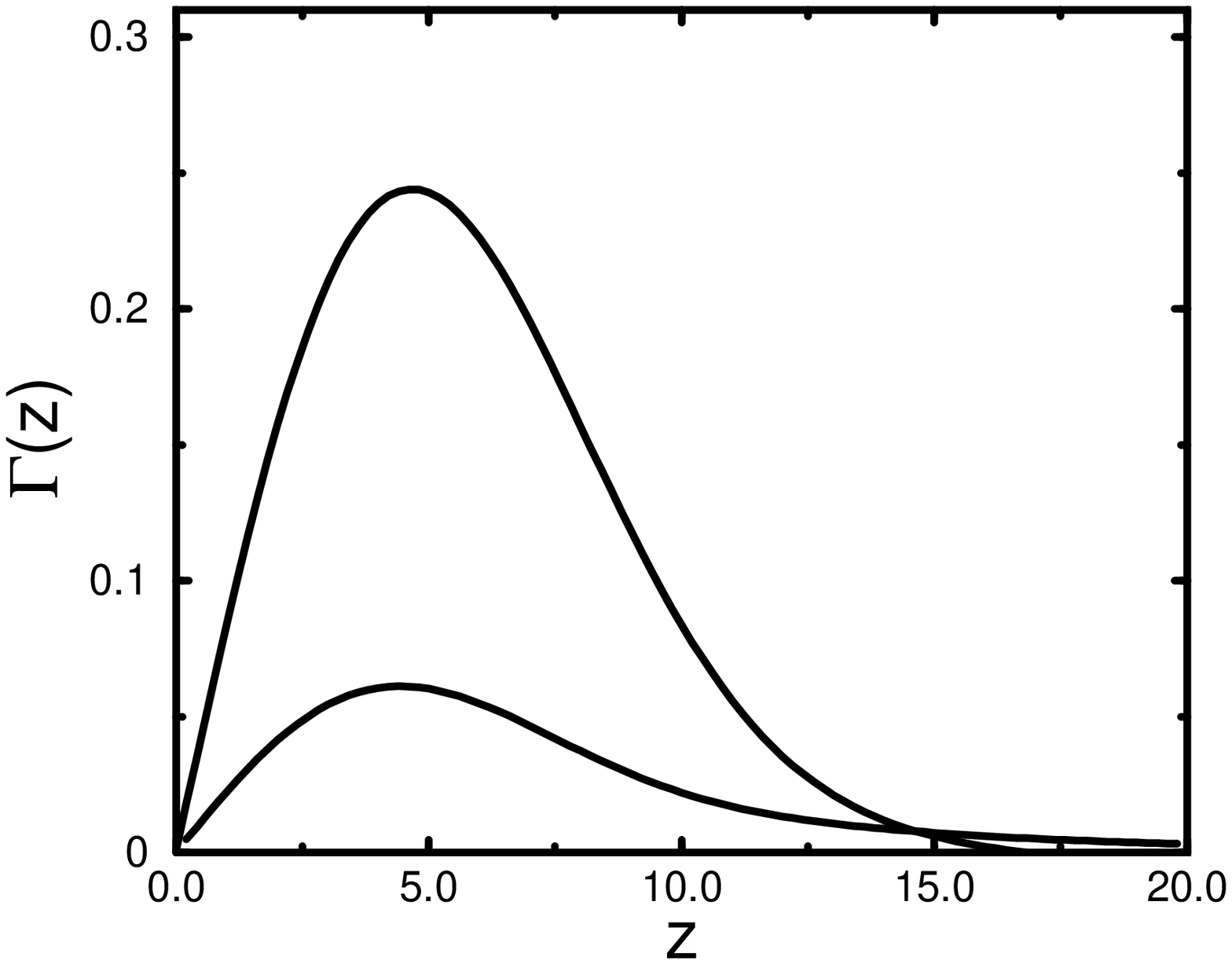}{13}
\label{fig_Ioffe_shape}
\end{minipage}
\begin{description}
  
\item[Fig.~4:] 
The Ioffe-time distributions corresponding to 
the polarized gluon distributions of Chiappetta et al. \cite{Chiap}
(upper curve) and and Brodsky et al. \cite{Bro95} (lower curve).  
\end{description}
\pagebreak


\begin{minipage}{16cm}
\bild{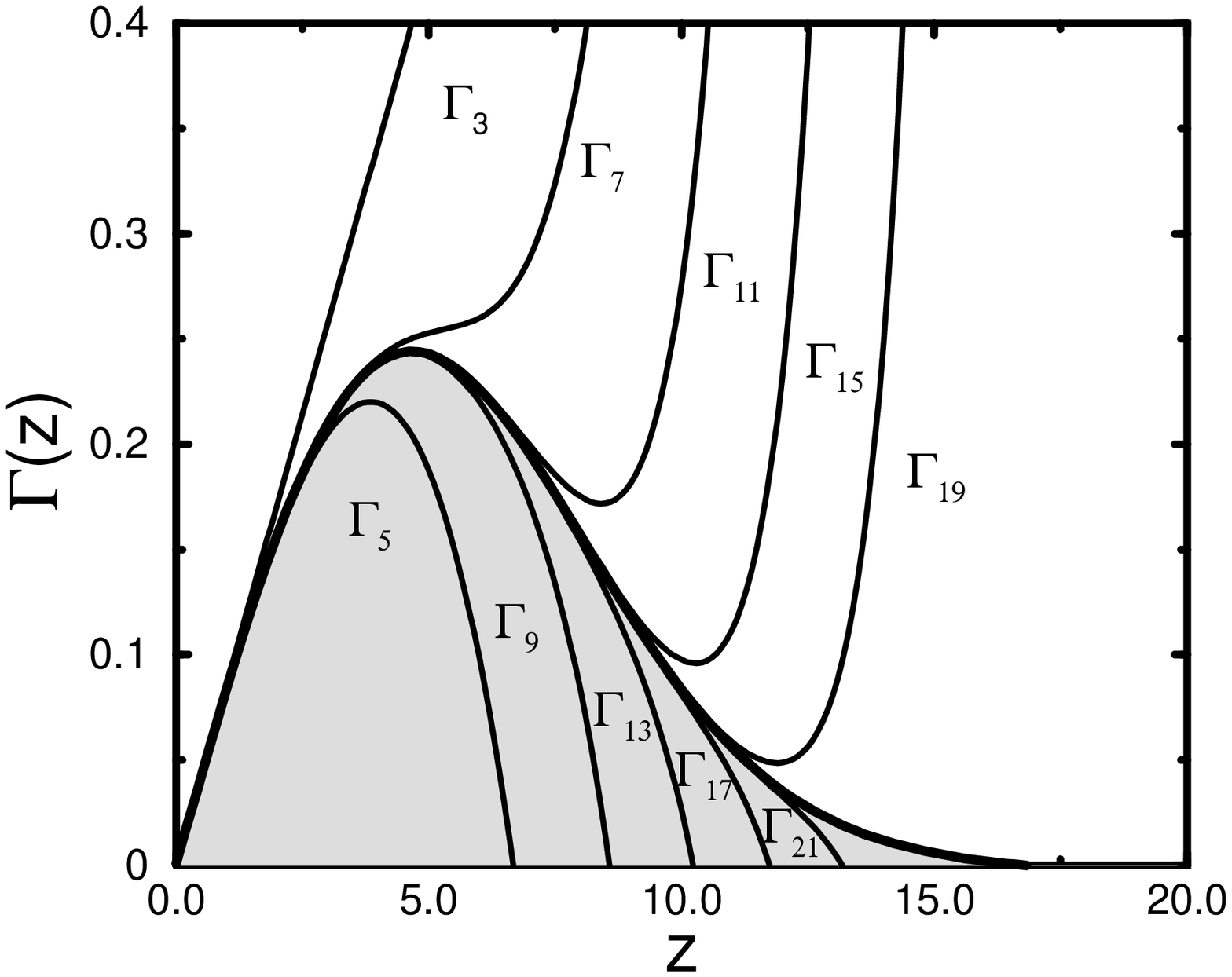}{13}
\label{fig_Ioffe_moments}
\end{minipage}
\begin{description}
  
\item[Fig.~5:]
  The Ioffe-time distribution of polarized glue, approximated by its 
  $n$-th order Taylor expansion around $z=0$ (\ref{eq:Gamma_expand}), 
  using the parameterization of Ref.\cite{Chiap}.
\end{description}
\pagebreak


\begin{minipage}{16cm}
\bild{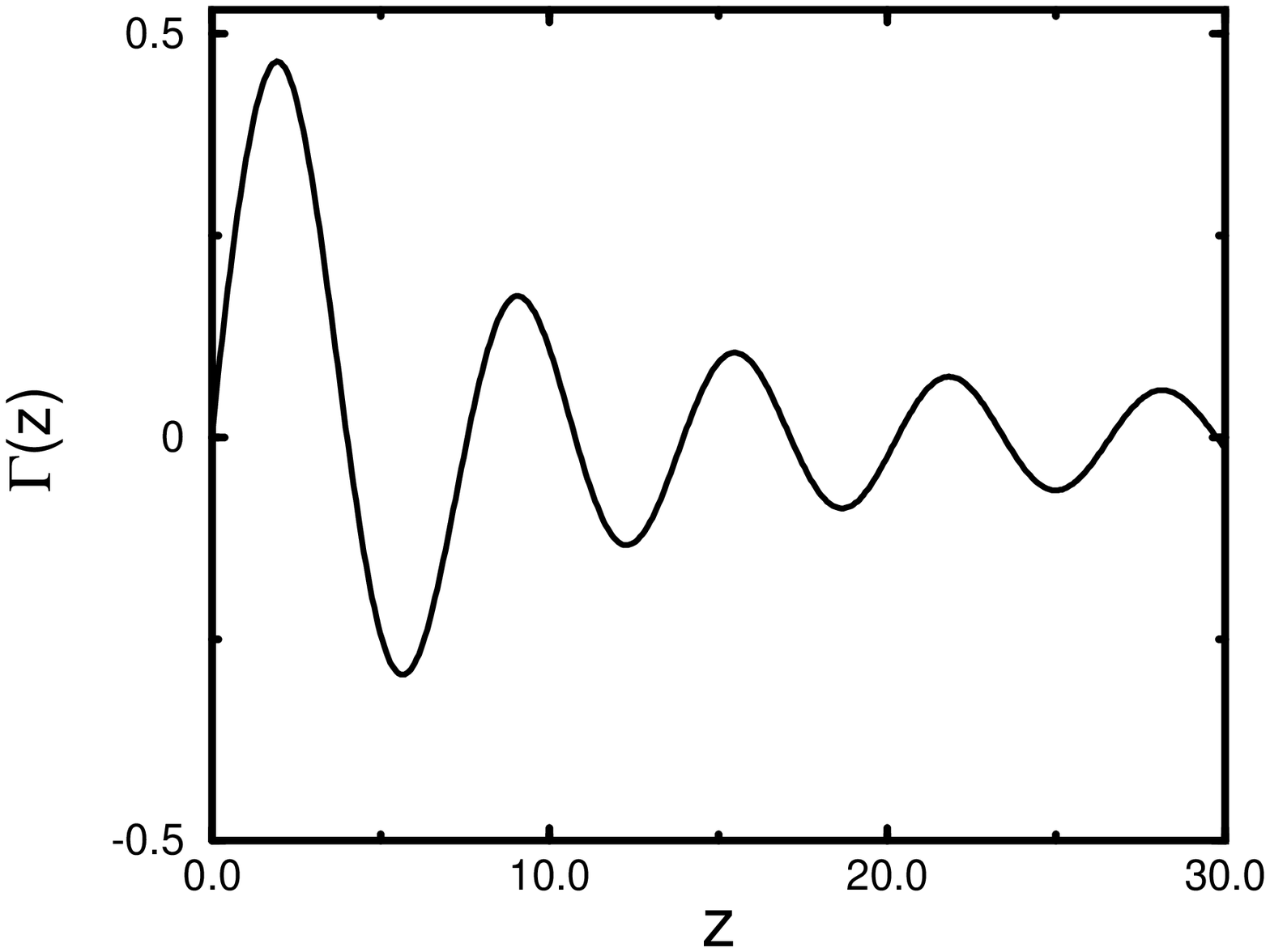}{13}
\end{minipage}
\begin{description}
  
\item[Fig.~6:] 
The Ioffe-time distribution $\Gamma(z)$ of polarized glue from 
QCD sum rules for the leading dimension-6 contribution,  
taken in the limit  $Q^2 \to 0$.
\end{description}


\begin{thebibliography}{99}

\bibitem{Fact}
J.C. Collins, D.E. Soper and G. Sterman, in {\em ``Perturbative
Quantum Chromodynamics''}, A.H. Mueller, ed. (World Scientific,
Singapore, 1989).

\bibitem{CERN}
   J. Ashman et al., Phys.~Lett.~{\bf B206}, 364 (1988);\\
   J. Ashman et al., Nucl.~Phys. {\bf B238}, 1 (1989);\\
   B. Adeva et al., Phys.~Lett. {\bf B302}, 500 (1993); \\
   B. Adeva et al., Phys.~Lett. {\bf B320}, 400 (1994); \\
   D. Adams et al., Phys.~Lett. {\bf B329}, 399 (1994);\\
   D. Adams et al., Phys.~Lett. {\bf B357}, 248 (1995).

\bibitem{SLAC}
 P.L. Anthony et al., Phys.~Rev.~Lett.~{\bf 71}, 959 (1993); \\
 K.~Abe et al., Phys.~Rev.~Lett. {\bf 74}, 346 (1995); {\it ibid}
   {\bf 75}, 25 (1995);\\
   K.~Abe et al., Phys.~Lett.~{\bf B364}, 61 (1995).
 
\bibitem{Deb96} 
  see e.g., 
  J.~Ellis, {\em Problems and Prospects in Spin Physics},
  hep-ph/9611208 (1996); \\
  S.~Forte, {\em Polarized Structure Functions: a Status Report},
  hep-ph/9610238
  (1996); \\
  X. Ji, {\em Hunting for the Remaining Spin in the Nucleon}, hep-ph/9610369
  (1996).
  
\bibitem{COMPASS} 
  G. Baum et al., {\em COMPASS proposal},
  CERN-SPSLC-96-14, 1996;\\
  R.G. Arnold et al., SLAC-Proposal-E156, 1997.
  
\bibitem{RHIC} 
  M. Bodo et al., 
  {\em Proposal on spin physics
   using the RHIC polarized collider}, BNL, Upton NY, 1992;\\
  S.~Guellenstern, P.~Gornicki, L.~Mankiewicz and 
  A.~Sch\"afer, Nucl.~Phys.~{\bf
    A560}, 494 (1993).

\bibitem{Alt96}
    G.~Altarelli, R.D.~Ball, S.~Forte and G.~Ridolfi, {\em Determination of the
    Bjorken Sum and Strong Coupling from Polarized Structure Functions},
    hep-ph/9701289 (1997).

\bibitem{Bag}
  R.L. Jaffe, Phys.~Rev.~{\bf D11}, 1953 (1975);\\
  F.E. Close and A.W. Thomas, Phys.~Lett.~{\bf B212}, 227 (1988);\\
  V.~Sanjose and V.~Vento, Phys.~Lett.~{\bf B225}, 15 (1988); Nucl.~Phys.~{\bf
    A501}, 672 (1989);\\
  A.W. Schreiber, A.I. Signal and A.W. Thomas, Phys.~Rev.~{\bf D44}, 2653  
  (1991);\\
  F.M. Steffens and A.W. Thomas, Prog.~Theor.~Phys.~Suppl.~{\bf 120}, 145
  (1995), and references therein.
  
\bibitem{Inst} D.~Diakonov, V.~Petrov, P.~Pobylitsa, M.~Polyakov  
  and  C.~Weiss,
  Nucl.~Phys.~{\bf B480}, 341 (1996).

\bibitem{QCDSR}
  A.V. Kolesnichenko, Sov.~J.~Nucl.~Phys.~{\bf 39}, 968 (1984);\\
  V.M. Belyaev and B.Yu. Blok, Z.~Phys.~{\bf C30}, 279 (1986);\\
  B.L. Ioffe and V.M. Belyaev, Nucl.~Phys.~{\bf B310}, 548 (1988).
  
\bibitem{Latt}
  M.~Fukugita et al., Phys.~Rev.~Lett.~{\bf 75}, 2092 (1995);\\
  S.J.~Dong et al., Phys.~Rev.~Lett.~{\bf 75}, 2096 (1995);\\
  M.~G{\"o}ckeler et al., Nucl.~Phys.~{\bf B42} (Proc.~Suppl.), 337 (1995); \\
  M.~G{\"o}ckeler et al., Phys.~Rev.~{\bf D53}, 2317 (1996);\\
  M.~G{\"o}ckeler et al., Nucl.~Phys.~{\bf B49} (Proc.Suppl), 250 (1996).
  
\bibitem{Collins}
  J.C. Collins and D.E. Soper, Nucl.~Phys.~{\bf B194}, 445 (1982); \\
  I.I. Balitskii and V.M. Braun, Nucl.~Phys.~{\bf B311}, 541 (1988/89).
  
\bibitem{Weig96}
  L. Mankiewicz and T. Weigl, Phys.~Lett.~{\bf B380}, 134 (1996); \\
  T.~Weigl and L.~Mankiewicz, Phys.~Lett.~{\bf B389}, 334 (1996).
  
\bibitem{Man90} A.~Manohar, Phys.~Rev.~Lett.~{\bf 65}, 2511 (1990).

\bibitem{BB91} I.I.~Balitskii and V.M.~Braun, Phys. Lett. {\bf B267}, 405
  (1991).

\bibitem{MPS97} L.~Mankiewicz, G.~Piller and 
  A.~Saalfeld, Phys.~Lett.~{\bf B395},
  318 (1997).

\bibitem{Jaffe96} R.L.~Jaffe, Phys. Lett. {\bf B365}, 359 (1996).
   
 \bibitem{Alt81} G.~Altarelli, Phys.~Rep.~{\bf 81}, 1 (1982).
   
 \bibitem{BGR97} J.~Bl\"umlein, B.~Geyer and D.~Robaschik, 
     {\em On the Evolution
     Kernels of Twist-2 Light Ray Operator for Unpolarized and Polarized Deep
     Inelastic Scattering}, hep-ph/9705264, (1997).
   
 \bibitem{BR97} I.I.~Balitskii, A.V.~Radyushkin, {\em Light Ray Evolution
     Equations and Leading Twist Parton Helicity Dependent Nonforward
     Distributions}, hep-ph/9706410 (1997).
   
 \bibitem{Ratc89} P.G.~Ratcliffe, Phys.~Lett.~{\bf B192}, 180 (1987);\\
   G.~Altarelli, G.G.~Ross, Phys.~Lett.~{\bf B212}, 391 (1988).
     
\bibitem{BGMS93} V.M.~Braun, P.~G{\'o}rnicki, L.~Mankiewicz and
     A.~Sch\"afer, Phys. Lett. {\bf B302}, 291 (1993).
     
\bibitem{Ste95} E.~Stein, P.~G{\'o}rnicki, L.~Mankiewicz and A.~Sch\"afer,
     Phys. Lett. {\bf B353}, 107 (1995); Phys. Lett. {\bf B343}, 369 (1995).

\bibitem{Bel96} A.V. Belitskii and O.V. Teryaev, Phys. Lett. {\bf B366}, 345
  (1996).

\bibitem{many} B.L. Ioffe and A.V. Smilga, Nucl. Phys. {\bf B216}, 373 (1983);
  \hfill\break V.A. Nesterenko and A.V. Radyushkin, Phys. Lett. {\bf B115}, 410
  (1982).

\bibitem{Regge} M.~Anselmino, A.~Efremov and E.~Leader, Phys.~Rep.~{\bf 261},
  1 (1995);
  
\bibitem{Bro95} S.J.~Brodsky, M.~Burkardt and I.~Schmidt, Nucl. Phys. {\bf
    B441}, 197 (1995).
  
\bibitem{Lad96} G.A.~Ladinsky, {\em A Collection of Polarized Parton
    Densities}, MSU-51120, hep-ph/9601287 (1996).
  
\bibitem{Chiap} P.~Chiappetta, P.~Colangelo, J.-Ph.~Guillet and G.~Narduli, Z.
  Phys.  {\bf C59}, 629 (1993).
  
\bibitem{GRV} M.~Gl\"uck, E.~Reya and A.~Vogt, Z. Phys. {\bf C67}, 433 (1995).
  
\bibitem{CTEQ} H.L.~Lai et al., {\em Improved Parton Distributions from Global
    Analysis of Recent Deep Inelastic Scattering and Inclusive Jet Data},
  hep-ph/9606399 (1996).
  
\bibitem{BGM95} V.M. Braun, P. G\'ornicki and L. Mankiewicz, Phys. Rev. {\bf
    D51}, 6036 (1995).
  
\bibitem{Ross96} G.G.~Ross and N.~Chamoun, Phys. Lett. {\bf B380}, 151 (1996).
  
\bibitem{Tony95}
  F.M.~Steffens, A.W.~Thomas, Nucl.~Phys.~{\bf A568}, 798 (1994);\\
  F.M.~Steffens, H.~Holtmann and 
  A.W.~Thomas, Phys.~Lett.~{\bf B358}, 139 (1995).


\end{thebibliography}
\end{document}